\documentclass[a4paper,11pt]{article}                    

\usepackage{soul}

\usepackage{graphicx}
\usepackage{comment}
\usepackage{amsmath,amssymb,amsbsy,latexsym}
\usepackage{amsmath}
\usepackage{color}
\usepackage{caption}
\usepackage{blindtext}

\usepackage{algorithm}
\usepackage[noend]{algpseudocode}

\usepackage[left=2.5cm,right=2.5cm,top=2.5cm]{geometry}

\newcommand{\mat}{\mathbf}
\newcommand{\onevec}{\mat 1}
\newcommand{\zerovec}{\mat 0}

\newcommand{\E}{\mathbb E}

\renewcommand{\P}{\mathbb P}
\newtheorem{theorem}{Theorem}
\newtheorem{lemma}{Lemma}
\newtheorem{remark}{Remark}
\begin{document}

\title{Mean Field Analysis of Personalized PageRank\\
with Implications for Local Graph Clustering}

\author{Konstantin Avrachenkov\thanks{Inria Sophia Antipolis, France}
\and Arun Kadavankandy\thanks{CentraleSup\'{e}lec, France}
\and Nelly Litvak\thanks{University of Twente, The Netherlands}}

\date{}

\maketitle

\begin{abstract}
We analyse a mean-field model of Personalized PageRank on the Erd\H{o}s-R\'{e}nyi random graph
containing a denser planted Erd\H{o}s-R\'{e}nyi subgraph. We investigate the regimes where the values
of Personalized PageRank concentrate around the mean-field value. We also study the
optimization of the damping factor, the only parameter in Personalized PageRank.
Our theoretical results help to understand the applicability of Personalized PageRank and
its limitations for local graph clustering.

\medskip

\end{abstract}

\section{Introduction}
\label{sec:intro}

Personalized PageRank (PPR) can be described as a random walk on a weighted graph. With probability $\alpha$ the random walk chooses one of the neighbour nodes, with probabilities
proportional to the edge weights, and with the complementary probability $1-\alpha$ the random walk
restarts from a set of seed nodes \cite{H03}. The analytical study of Personalized
PageRank is challenging because of its non-local nature. Interestingly enough, it
is easier to analyse Personalized PageRank on directed graphs. In \cite{AL06},
the expected values of the standard PageRank \cite{Petal99} with uniform restart and
Personalized PageRank have been analyzed for directed preferential attachment graphs.
In \cite{VL10} a stochastic recursive equation has been derived for
the Personalized PageRank on directed configuration model. This equation has been
thoroughly analyzed in \cite{Cetal17,JO10} and in the works mentioned therein.

On the other hand, the analysis of Personalized PageRank on undirected random graph models is more
difficult because a simple random walk on an undirected graph can pass through an edge in both directions,
thus creating many short cycles and loops. To the best of our knowledge,
\cite{Aetal17} is the only work studying Personalized PageRank on undirected
Erd\H{o}s-R\'{e}nyi (ER) random graphs and stochastic block models. For the
analysis of \cite{Aetal17} to hold, the personalization vector or the restart
distribution has to be sufficiently delocalized. In \cite{Fetal07} a mean-field
model for the standard PageRank has been proposed without a formal justification.
In the recent work \cite{KUK16} a mean-field model has been proposed for a modification
of Personalized PageRank where the contributions from all paths are same.
The authors of \cite{KUK16} have carried out their analysis in dense stochastic
block models when the edge probabilities are fixed, i.e., they do not scale with the size of the graph.

In the present work we analyze Personalized PageRank with a localized restart
distribution. As a graph model, we consider an ER random graph with a smaller denser
ER graph planted within. We establish conditions for concentration and non-concentration of PPR
under different scaling laws of the edge probabilities.
In particular, we show that when the graph is not too sparse there is a concentration to the mean field model of PPR when
the size of subgraph scales linearly and the number of seeds scales sufficiently fast with the graph size.
In other words, we establish sufficient conditions for the convergence of PPR to its mean-field form in medium dense graphs.
In addition, we show that these conditions are also necessary in a class of sparse graphs with tree-like local
structure; i.e., if the number of seed nodes is too small, PPR does not concentrate in a class of sparse graphs.
We also show that when there is concentration, the values of PPR can be well approximated
by a simple mean-field model and this model can be used for instance for the optimal
choice of the damping factor.

An ER graph with a planted denser ER subgraph is the simplest model of a random graph
with heterogeneity.
It is also a good benchmark model for testing various local graph
clustering algorithms. Local graph clustering algorithms are gaining
importance since often in practice one would like to recover one particular
cluster of a graph using as a guide a few representative seed nodes. One of the
first efficient local clustering algorithms is the Nibble algorithm \cite{ST04,ST13}
with quasi-linear complexity. The Nibble algorithm is truncation based
approximation of a few steps of a lazy random walk.
In \cite{ACL06,ALa06} a modification of the Nibble algorithm
using Approximate Personalized PageRank (APPR) has been proposed and evaluated.
APPR has lighter computational complexity than the Nibble algorithm.
In \cite{GM14} it has been shown that APPR can be obtained as a solution of an
optimization problem with $l_1$-regularization, which ensures sparsity in APPR
elements. Both Nibble and APPR try to keep the probability mass localized on
few important elements.
Recently, in \cite{Aetal16} a further improvement to the Nibble algorithm has been
proposed based on the technique of the evolving sets.

Our results imply that one needs a significant number of seed nodes to
obtain a high quality local cluster. If there are only a few seed nodes, both
PPR and APPR suffer from non-concentration. Specifically, the main reason for
the non-concentration of PPR and APPR is the significant leakage of probability
mass via the seed nodes' neighbours which are outside of the target community.

The methods in \cite{ST04,ST13,ACL06,ALa06,Aetal16} aim to find a local cut with
target conductance. However, in \cite{CKL17} and \cite{ZLM13} significant limitations
of the random walks and PPR based local clustering methods are presented in terms of graph
conductance and related quantities. As a by-product of our analysis, in Subsection~\ref{subsec:cond}
we show that the natural cluster in our random graph model also does not really
correspond to the problem of conductance minimization. This observation can be viewed
as complementary to the results in \cite{CKL17} and \cite{ZLM13}.


We would like to note that in \cite{HWX16} semidefinite relaxation is used
to recover a hidden subgraph without seed nodes
and in \cite{Ketal17} the belief propagation based algorithm is used to recover
a subgraph with seed nodes. These two methods appear to be superior to Personalized
PageRank based methods on the considered random graph models but require graph parameters
as an input. It is interesting to observe that in the semi-supervised clustering
\cite{AVG10,ZMZ14} the detectability transition disappears when any linear fraction
of seed nodes is introduced.

This paper is organized as follows. In the following section we formally define
the random graph model and describe the mean field approximation of Personalized PageRank.
In Section~\ref{sec:concentration} we show that there
is a concentration in the mean field model of PPR when the size of subgraph and
the number of seeds scale linearly with the graph size. However, as demonstrated
in Section~\ref{sec:nonconcentration}, if the number of seed nodes is too small,
there is no concentration. Then, in Section~\ref{sec:dampingopt} using the mean field model
we provide a recommendation for setting the restart probability. Section~\ref{sec:localcluster}
concludes the technical part of the paper with numerical illustrations and discussions
about possible limitations of the PPR and APPR based local graph clustering methods.
Finally, in Section~\ref{sec:conclusions} we recall our main results and outline
promising avenues for further research.

\section{Graph model and notations}
\label{sec:model}

In this section we introduce the model and notations. The notations are summarised in Table~\ref{tab:notation} at the end of the paper.

We consider an ER random graph $G(n,p)$ with a planted ER subgraph $G(m,q)$.
We are interested in the case when the planted ER subgraph is denser than
the background ER graph, i.e., when $q > p$.
Without loss of generality, we assume that the indices of the subgraph nodes
coincide with the first $m$ indices of the background graph $G(n,p)$.
Denote this set of vertices corresponding to the planted subgraph by $\mathcal C.$
At the moment we do not specify any scaling for $m$, $p$ and $q$ and shall
discuss various scalings whenever it is needed. We denote by $A=\{a_{ij}\}$ the adjacency
matrix of the resulting graph, i.e.,
$$
a_{ij} =
\left\{\begin{array}{ll}
1, & \mbox{if $i$ is a neighbour of $j$,}\\
0, & \mbox{otherwise}.
\end{array}\right.
$$
Denote by $\onevec_n$ the column vector of ones of dimension $n$ and by $\mat{J}_{m,n}$ the matrix
of ones of dimensions $m$-by-$n$. Also denote by $\zerovec_n$ the column vector of zeros
of dimension $n$. Let $d=A\onevec_n$ be the vector of nodes' degrees
and $D=diag\{d\}$ is the diagonal matrix of nodes' degrees. Then, $P=D^{-1}A$ is
the transition probability matrix of the standard random walk when the walker chooses
the next node to visit uniformly among the neighbours of the current node.
Let $k$ nodes of the planted subgraph be disclosed to us. Of course, $k \le m$. Again without the loss
of generality, we can assume that these $k$ seed nodes correspond to the first
$k$ nodes of the background graph. Denote the set of seed nodes by $\mathcal S.$
Then the personalization vector or the restart distribution $\nu$ is given by
$$
\nu = \left[ \frac{1}{k}\onevec_k^T \ \zerovec_{n-k}^T \right].
$$
Personalized PageRank $\pi$ can be expressed as follows:
\begin{equation}
\label{eq:PPR}
\pi = (1-\alpha) \nu [I-\alpha P]^{-1}.
\end{equation}

Now let us define the mean field model of Personalized PageRank. It is based on the
expected adjacency matrix:
$$
\bar{A} =
\left[\begin{array}{cc}
q\mat{J}_{m,m} & p\mat{J}_{m,n-m} \\
p\mat{J}_{n-m,m} & p\mat{J}_{n-m,n-m}
\end{array}\right],
$$
and the associated mean field transition probability matrix $\bar{P}=\bar{D}^{-1}\bar{A}$.
The mean field Personalized PageRank is given by
\begin{equation}
\label{eq:MFPPR}
\bar{\pi} = (1-\alpha) \nu [I-\alpha \bar{P}]^{-1}.
\end{equation}
Note that due to symmetry, the mean field Personalized PageRank has the following structure
$$
\bar{\pi} = \left[ \bar{\pi}_0\onevec_k^T \ \bar{\pi}_1\onevec_{m-k}^T \ \bar{\pi}_2\onevec_{n-m}^T \right],
$$
where $\bar{\pi}_i, i=0,1,2,$ are determined by the system of linear equations:
\begin{equation}
\label{eq:barpi0s}
\bar{\pi}_0 - \bar{\pi}_0 \frac{\alpha k q}{mq + (n-m)p} - \bar{\pi}_1 \frac{\alpha (m-k) q}{mq+(n-m)p}
- \bar{\pi}_2 \frac{\alpha (n-m)}{n} = \frac{1-\alpha}{k},
\end{equation}
\begin{equation}
\label{eq:barpi1s}
- \bar{\pi}_0 \frac{\alpha k q}{mq+(n-m)p} + \bar{\pi}_1 - \bar{\pi}_1 \frac{\alpha (m-k) q}{mq+(n-m)p}
- \bar{\pi}_2 \frac{\alpha (n-m)}{n}= 0,
\end{equation}
\begin{equation}
\label{eq:barpi2s}
- \bar{\pi}_0 \frac{\alpha k p}{mq+(n-m)p} - \bar{\pi}_1 \frac{\alpha (m-k) p}{mq+(n-m)p}
+ \bar{\pi}_2 - \bar{\pi}_2 \frac{\alpha (n-m)}{n}= 0.
\end{equation}
These equations can easily be solved in explicit form. For instance, subtracting equation (\ref{eq:barpi1s})
from equation (\ref{eq:barpi0s}) we obtain
\begin{equation}
\label{eq:barpi01}
\bar{\pi}_0 = \frac{1-\alpha}{k} + \bar{\pi}_1.
\end{equation}
Multiplying equation (\ref{eq:barpi1s}) by $p$ and equation (\ref{eq:barpi2s}) by $q$, respectively, and then
subtracting one from another, we get
\begin{equation}
\label{eq:barpi12}
\bar{\pi}_1 = \bar\pi_2 \left( \frac{q}{p} - \frac{\alpha(n-m)}{n}\frac{q}{p} + \frac{\alpha(n-m)}{n} \right).
\end{equation}
Then, substituting subsequently (\ref{eq:barpi01}) and (\ref{eq:barpi12}) into (\ref{eq:barpi2s}) yields
\begin{equation}
\label{eq:barpi2}
\bar\pi_2 = \frac{(1-\alpha) \alpha p}
{(mq+(n-m)p)\left(1-\frac{\alpha(n-m)}{n}\right)
- \alpha m \left(q-\frac{\alpha(n-m)}{n}(q-p)\right)}.
\end{equation}
Using (\ref{eq:barpi01}) and (\ref{eq:barpi12}), one easily retrieves $\bar\pi_1$ and $\bar\pi_2$.
Namely, we have
\begin{equation}
\label{eq:barpi1}
\bar\pi_1 = \frac{(1-\alpha) \alpha \left(q-\frac{\alpha(n-m)}{n}(q-p)\right)}
{(mq+(n-m)p)\left(1-\frac{\alpha(n-m)}{n}\right)
- \alpha m \left(q-\frac{\alpha(n-m)}{n}(q-p)\right)},
\end{equation}
and
\begin{equation}
\label{eq:barpi0}
\bar\pi_0 = \frac{1-\alpha}{k} + \frac{(1-\alpha) \alpha \left(q-\frac{\alpha(n-m)}{n}(q-p)\right)}
{(mq+(n-m)p)\left(1-\frac{\alpha(n-m)}{n}\right)
- \alpha m \left(q-\frac{\alpha(n-m)}{n}(q-p)\right)}.
\end{equation}
We would like to note that there is a simple bound on the expectated value of PPR.
Clearly, we have $\sum_{i\in\mathcal C \backslash \mathcal S}\pi_i \le 1.$
By taking expectation of both sides and using symmetry, we obtain
\begin{equation}
\label{eq:exprank}
\mathbb E(\pi_i) \le \frac{1}{m-k},
\end{equation}
for $i\in \mathcal C \backslash \mathcal S.$ Similarly, we have
\begin{equation}
\label{eq:exprank1}
\mathbb E(\pi_i) \le \frac{1}{n-m},
\end{equation}
for $i\in \{1,2,\ldots,n\} \backslash \mathcal C.$

\section{Conditions for concentration of the PPR}
\label{sec:concentration}

Let us study the conditions when Personalized PageRank concentrates around its mean-field model.
In order to investigate different regimes, we shall emphasize the dependence of the key parameters
on the size of the graph $n$, that is $k:=k(n)$, $m:=m(n)$, $p:=p(n)$ and $q:=q(n)$.
The following result states the $L^2$ convergence of the relative error of the mean field Personalized PageRank.

\begin{theorem}\label{thm:L2conv}
Assume that $nq(n)=\omega(\log(n))$ and $p(n)/q(n)=\Theta(1)$.
Then, the relative $L^2$ distance between $\pi$ and $\bar{\pi}$ converges in probability to zero.
More precisely, there exists $C>0$ such that
\begin{equation}
\label{eq:bound_final}
\frac{||\pi-\bar{\pi}||_2}{||\bar{\pi}||_2} \le \frac{\alpha C}{(1-\alpha)\sqrt{\frac{np(n)}{\log(n)}} - \alpha C}, \quad \mbox{a.a.s.}
\end{equation}
\end{theorem}
\noindent {\bf Proof:}
It follows from the sensitivity analysis of the system of linear equations \cite{GV13},
$(A+\Delta A)(x +\Delta x)=b$, that the following inequality takes place
\begin{equation}
\label{eq:sensbound}
\frac{||\Delta x||_2}{||x||_2} \le \frac{||A^{-1}||_2||\Delta A||_2}{1-||A^{-1}||_2||\Delta A||_2}.
\end{equation}
In our case, this general inequality becomes
\begin{equation}
\label{eq:bound_specific}
\frac{||\pi-\bar{\pi}||_2}{||\bar{\pi}||_2} \le \frac{||[I-\alpha \bar{P}]^{-1}||_2||-\alpha[P-\bar{P}]||_2}
{1-||[I-\alpha \bar{P}]^{-1}||_2||-\alpha[P-\bar{P}]||_2}.
\end{equation}
Since one is the maximal modulus eigenvalue of $\bar{P}$, we have
\begin{equation}
\label{eq:boundinv}
||[I-\alpha \bar{P}]^{-1}||_2 = \frac{1}{1-\alpha}.
\end{equation}
From Lemma~\ref{l:normbound}, which we provide below, it follows that there is $C>0$ such that
\begin{equation}
\label{eq:bounddelta}
||\alpha[ P-\bar{P}]||_2 \le \alpha C\sqrt{\frac{\log(n)}{np(n)}}, \quad \mbox{a.a.s.}
\end{equation}
The combination of (\ref{eq:bound_specific}), (\ref{eq:boundinv}) and (\ref{eq:bounddelta}) yields the result. \hfill $\Box$

We would like to note that the inequality (\ref{eq:bound_final}) indicates very slow convergence.
Indeed, if we consider the standard moderately sparse regime with
\begin{equation}
\label{eq:semidense}
p(n) = \frac{\log^c(n)}{n}, \quad c > 1,
\end{equation}
the rate of convergence will be of the order $1/\log^{\frac{c-1}{2}}(n)$.

We will now provide Lemma~\ref{l:normbound}, which is crucial for the proof of Theorem~\ref{thm:L2conv}.

\begin{lemma}
\label{l:normbound}
Assume that $nq(n)= \omega(\log(n))$, and $p(n)/q(n) = \Theta(1).$ Then for some $C>0$,
\[
|| P-\bar{ P}||_2 \le C \sqrt{\frac{\log(n)}{np}}, \quad \mbox{a.a.s.}
\]
\end{lemma}
\noindent {\bf Proof:}
  We denote by $\bar{ A}$ the expected adjacency matrix and by $\bar D$ the diagonal matrix of expected degrees. From Lemma~10 in \cite{Aetal17} we have
  \begin{equation}\label{eq_adj_conc}
  \| A - \bar{ A}\|_2 \le K\sqrt{\log(n)nq(n)}, \quad \mbox{a.a.s.},
\end{equation}
where we used the fact that $q(n)>p(n).$
Then, since $p(n)/q(n) = \Theta(1),$ we also have $np(n) = \omega(\log(n)).$ Therefore, from Lemma 8 of \cite{Aetal17}
for some $C_1>0$ we have
\begin{equation}\label{eq_deg_conc}
\|\bar{ D}^{-1} D -  I\|_2 \le C_1 \sqrt{\frac{\log(n)}{np(n)}}, \quad \mbox{a.a.s.},
\end{equation}
since $\bar{ A}$ is a rank two matrix with all entries in the upper-left sub-matrix of size $|\mathcal C|\times|\mathcal C|$ equal to $q(n)$ and all other entries being $p(n)$.
Now, from (\ref{eq_adj_conc}) we obtain
\begin{equation}\label{eq_adj_bound}
  \| A\|_{2} \le \| A - \bar{ A}\|_{2} + \|\bar{ A}\|_{2}
  \le K \sqrt{\log(n)nq(n)} + nq(n), \quad \mbox{a.a.s.}
\end{equation}
Using the above bounds, we get
\begin{align*}
  \| D^{-1} A - \bar { D}^{-1}\bar { A}\|_{2} &= \|( D^{-1}\bar { D} -  I)\bar { D}^{-1}{ A} +\bar { D}^{-1}( A-\bar{ A})\|_2\\
  &= \| D^{-1}\bar { D} - I\|_{2}\|\bar{ D}^{-1}\|_{2}\| A\|_{2}
  + \|\bar{ D}^{-1}\|_{2}\| A-\bar{ A}\|_{2}\\
  &\le\frac{1}{np(n)}\left( C_1 \sqrt{\frac{\log(n)}{np(n)}}( K\sqrt{\log(n)nq(n)} + nq(n))
  + K\sqrt{\log(n)nq(n)}\right)\\
  &\le C\sqrt{\frac{\log(n)}{np(n)}}
\end{align*}
for some $C>0$. \hfill $\Box$

Now let $U$ be a uniformly randomly sampled integer between 1 and $n$ and
\[\eta(i)=
\left\{\begin{array}{ll}
0, & 1 \le i \le k,\\
1, & k+1 \le i \le m,\\
2, & m+1 \le i \le n.
\end{array}\right.
\]
Then, using Theorem~\ref{thm:L2conv} we can establish that the difference between $\pi_U$ and $\bar\pi_{\eta(U)}$ is vanishing with high probability when $k(n)$ is large enough.
The result is formally stated in the next theorem.

\begin{theorem}\label{thm:concentration_U}
Let conditions of Theorem~\ref{thm:L2conv} hold. Furthermore, let $k=k(n)$ be such that $k(n)p(n)=\omega(\log(n))$ and let $U$ be the index of a randomly sampled node $1,2,\ldots,n$.
Then, for any $\varepsilon>0$ the following result holds
\begin{equation}
\label{eq:concetration_theorem}
\lim_{n\to\infty}\P\left(|\pi_U-\bar\pi_{\eta(U)}|\ge \varepsilon n^{-1}\right)=0.
\end{equation}
\end{theorem}
\noindent {\bf Proof:}
Denote by $B_n$ the event that the inequality in (\ref{eq:bound_final}) holds:
\[B_n=\left\{\frac{||\pi-\bar{\pi}||_2}{||\bar{\pi}||_2} \le \frac{\alpha C}{(1-\alpha)\sqrt{\frac{np(n)}{\log(n)}} - \alpha C}\right\}\]
for an appropriate value of $C$.
The idea of the proof is to use this inequality to bound our probability of interest on $B_n$ and then use the fact that  $\lim_{n\to\infty}P(B_n)=1$.

To this end, we need to bound the probability in (\ref{eq:concetration_theorem}) using $L^2$-norms. We do this by first conditioning on the realization of $G$. We will denote by $\P_n$ the probability measure conditioned on $G$. Then the randomness is only in
the choice of $U$. By Markov's inequality, we have
\begin{align*}
\P_n\left(|\pi_U-\bar\pi_{\eta(U)}|\ge \varepsilon n^{-1}\right)&\le\frac{n\E_n(|\pi_U-\bar\pi_{\eta(U)}|)}{\varepsilon}\\
&=\frac{n}{\varepsilon}\left(\frac{1}{n}\sum_i|\pi_i-\bar\pi_{\eta(i)}|\right)\\
&\le \frac{n}{\varepsilon}\left(\frac{1}{n}\sum_i|\pi_i-\bar\pi_{\eta(i)}|^{2}\right)^{1/2}\\
&= \frac{\sqrt{n}}{\varepsilon}\,||\pi-\bar\pi||_2.
\end{align*}
Next, by the full probability formula, we have
\begin{align}
\nonumber
\P\left(|\pi_U-\bar\pi_{\eta(U)}|\ge \varepsilon n^{-1}\right)&=\E\left[\P_n\left(|\pi_U-\bar\pi_{\eta(U)}|
\ge \varepsilon n^{-1}\right)\right]\\\nonumber
& = \E\left[P_n\left(|\pi_U-\bar\pi_{\eta(U)}|
\ge \varepsilon n^{-1}\right)\onevec\{B_n\}\right]+
\E\left[P_n\left(|\pi_U-\bar\pi_{\eta(U)}|\ge \varepsilon n^{-1}\right)\onevec\{\bar{B}_n\}\right]\\\nonumber
&\le \E\left[\frac{\sqrt{n}}{\varepsilon}\,||\pi-\bar\pi||_2\,\onevec\{B_n\}\right]+P(\bar{B}_n)\\
\label{eq:thm2_pf}
&\le \frac{\alpha C \sqrt{n}\, ||\bar\pi||_2}{(1-\alpha)\sqrt{\frac{np(n)}{\log(n)}} - \alpha C}\,P({B}_n)+P(\bar{B}_n),
\end{align}
where we used \eqref{eq:bound_final} for the last step. Since $P({B}_n)$ converges to one as $n\to\infty$, the statement of the theorem follows when $\frac{\alpha C \sqrt{n} ||\bar\pi||_2}{(1-\alpha)\sqrt{\frac{np(n)}{\log(n)}} - \alpha C}$ converges to zero. It remains to verify that this is indeed the case when $\frac{k(n)p(n)}{\log(n)}\to\infty$. Now note that  \eqref{eq:thm2_pf} together with the fact that $\bar\pi_0=\Theta((k(n))^{-1})$, $\bar\pi_1=\Theta((m-k(n))^{-1})$, $\bar\pi_2=\Theta((n-m)^{-1})$, and $k=O(m)$, implies that
\[
||\bar\pi||_2=\left(k(n)\bar\pi_0^2+(m-k(n))\bar\pi_1^2+(n-m)\bar\pi_2^2\right)^{1/2}
=\Theta\left(k(n)^{-1/2}\right),
\]
which gives the result. \hfill $\Box$

The practical implication of Theorem~\ref{thm:concentration_U} is that the condition $\frac{k(n)p(n)}{\log(n)}\to\infty$ is sufficient for $\pi$ to be well approximated by $\bar{\pi}$. In other words, $\pi$ is concentrated around $\bar{\pi}$. Notice that the result of Theorem~\ref{thm:concentration_U} holds for a large range of regimes. Indeed, the requirement $k(n)p(n)=\omega(\log(n))$ means that the number of seed node neighbours of a node $i\notin {\cal C}$ is of the order larger than $\log(n)$. This condition is satisfied
in a dense regime as well, when $p(n)$ and $q(n)$ are constants.
For example, the above analysis is also applicable in the setting of \cite{KUK16}, but without the artificial modification of PPR.
In the next section we will focus on the regimes where the local tree approximation of the graph is valid. This does not include a dense regime or any regime where $np(n)$, $nq(n)$ are powers of $n$. In this class of regimes, we will obtain conditions, under which concentration does not occur.


\section{Non-concentration conditions for PPR}
\label{sec:nonconcentration}

In this section we will, as before, assume that $p(n)/q(n)=\Theta(1)$ and consider the regime when $nq(n)$ is smaller
than a power of $n$, more precisely, $nq(n)=o(n^\varepsilon)$ for all $\varepsilon>0$. Note this includes our regime of interest (\ref{eq:semidense}). We will show that in this range of parameters the condition $k(n)q(n)\to\infty$ is necessary for concentration of $\pi$ around $\bar{\pi}$ to occur. Specifically, when this condition is violated, then $\pi$ does not concentrate at all, that is, the coefficient of variation of $\pi_i$, $i=1,2,\ldots,n$, is non-vanishing.

Our argument relies on the local tree approximation of our random graph model constructed as follows. For any node $i$  in $G$, we will say that $i$ is of type ${\cal C}$ if $i\in {\cal C}$ and of type $\bar{\cal C}$ otherwise. Consider a rooted Galton-Watson tree ${\cal T}_i^t$ of depth $t$ with root $i$. Assume that each node has $\text{Poisson}(mq(n))$ number of offspring of type ${\cal C}$ and $\text{Poisson}((n-m)p(n))$ number of offspring of type $\bar{\cal C}$, each type independent of each other. The following lemma from \cite{BWX15} states that ${\cal T}_i^t$ can be coupled with high probability with the $t$-hop neighborhood of a random node, $G_i^t$.

\begin{lemma}\cite[Lemma~10]{BWX15}
\label{l:coupling}
Assume that $p(n)/q(n)=\Theta(1)$ and $nq(n)=o(n^\varepsilon)$ for all $\varepsilon>0$. Then for any node $i=1,2,\ldots,n$ and $t:=t(n)\to\infty$ such that $(nq(n))^t = n^{o(1)},$ there exists a coupling such that $(G_i^t,\sigma^t) = ({\cal T}_i^t,\tau^t)$  with probability $1 - n^{-1 + o(1)},$ where $G_i^t$ is the subgraph induced by the set of nodes at distance $t$ from $i$ and $\sigma^t$ is the vector of
the types of the nodes of the graph. Also, ${\cal T}_i^t$ is a Galton-Watson tree with Poisson offspring
distribution and $\tau^t$ is the vector of types on ${\cal T}_i^t.$
\end{lemma}

We want to show that if $k(n)q(n)=O(1)$, then,  with positive probability,  the difference between $\pi_i$ and $E(\pi_i)$ is of the same order of magnitude as $\E(\pi_i)$ itself. The prove consists of two steps. First, in Lemma~\ref{l:finite_neighborhood} we will show that $\pi_i$ is well approximated by a $t$-neighborhood. Then, in Theorem~\ref{thm:nonconPPR}, we will use this result together with Lemma~\ref{l:coupling} to demonstrate the non-concentration.

Denote by $\pi^t$ the contribution of paths shorter than $t$:
\begin{align*}
\pi^t&=(1-\alpha) \nu \sum_{l=0}^{t-1}\alpha^l P^l.
\end{align*}
Now we can easily prove Lemma~\ref{l:finite_neighborhood} below.

\begin{lemma}
\label{l:finite_neighborhood} Take $t:=t(n)\to\infty$. Then for any $i\notin {\cal S}$,
\begin{equation}
\label{eq:finite_neighborhood_E}
\E(|\pi_i-\pi_i^t|)=o(n^{-1}).
\end{equation}
\end{lemma}

\noindent {\bf Proof:}
First, we split the PPR in (\ref{eq:PPR}) as follows:
\[\pi = \pi^t + (1-\alpha)\nu\alpha^tP^t\sum_{l=0}^{\infty}\alpha^l P^l.\]
Now, proceeding exactly as in \cite{Cetal17}, for the second term we get
\begin{align*}
\|\pi_i-\pi_i^t\|_1 &=\alpha^t.
\end{align*}
Assume that $i\in {\cal C}\backslash {\cal S}$, and note that for the nodes outside ${\cal C}$ the argument is exactly the same. Since all $m$ nodes in $\cal C$ are symmetric, for any $t=t(n)\to\infty$ we immediately obtain
\[\E(|\pi_i-\pi_i^t|)\le \frac{1}{m}\,\alpha^t=o(n^{-1}).\]
This gives (\ref{eq:finite_neighborhood_E}). \hfill $\Box$


Now using Lemma's~\ref{l:coupling} and \ref{l:finite_neighborhood} we can prove the non-concentration result, stated in the next theorem. We will prove the result in the regime when $k(n)$ is at least a power of $n$ (however, such power can be arbitrarily small).

\begin{theorem}(Non-concentration of PPR)
\label{thm:nonconPPR}
Let $G$ be rooted at node $i\notin {\cal S}$. If $k(n)q(n) = O(1)$ and there exists $\xi>0$ such that $k(n)\ge n^\xi$, then
\begin{equation}
\label{eq:non-concenration}
\frac{\mbox{Var}(\pi_i)}{\E^2(\pi_i)}=\Omega(1).
\end{equation}
\end{theorem}

\noindent {\bf Proof:}
We will again prove the result for node $i\in {\cal C}\backslash {\cal S}$.  As in Lemma~\ref{l:finite_neighborhood}, the argument for $i\notin {\cal C}$ is exactly the same. First of all, note that
\[\E(\pi_i) \le \frac{1}{m}= \Theta(n^{-1})\]
because $\E(\pi_i)\le \frac{1}{m}=\Theta(n^{-1})$. Next, taking into account only neighbors of $i$ from ${\cal S}$ and using Jensen's inequality, we can write
\begin{align*}
\E(\pi_i)&\ge \frac{\alpha}{k(n)}\,\E\left(\sum_{j\in {\cal S}}\frac{1\{a_{i,j}=1\}}{d_j}\right)\\&
= \frac{\alpha}{k(n)}\, k(n) q(n)\,\E\left(\left.\frac{1}{d_j}\right| a_{i,j}=1\right)\\&
\ge\frac{\alpha q(n)}{\E(d_j|a_{i,j}=1)}\\&
= \frac{\alpha q(n)}{((m-1)q(n)+np(n)+1)}= \Theta(n^{-1}),
\end{align*}
where we recall that $a_{i,j}$ is the element of the adjacency matrix $A$. In the rest of the proof we will evaluate $\mbox{Var}(\pi_i)$. For that, we will use the decomposition of PageRank from \cite{NK06}.  Consider a simple random walk $(X_l)_{l\ge 0}$ on $G$ such that at each step the walk continues with probability $\alpha$ and terminates with probability $1-\alpha$. Let $T$ be the termination time, which has a geometric distribution with parameter $(1-\alpha)$. Denote by $\P_{(j)}$ the conditional probability given the event $\{X_0=j\}$. Then, for any realization of the graph, from \cite{NK06} we have:
  \begin{align}
  \label{eq:decomposition}
  {\pi}_i&=\frac{(1-\alpha)}{|{\cal S}|}\,\left(\sum_{s\in {\cal S}}\,\P_{(s)}(\mbox{$X_l=i$ for some $l\le T$})\right)\mathbb{E}_{(i)}\left[\sum_{l=0}^{\infty}1\{X_l=i\}\right].\\
  &=(I)\times (II)\times (III).\nonumber
  \end{align}
   Here (I), (II) and (III) are random variables that depend on a realization of the random graph. Note that (III) is the average number of visits to $i$, starting from $i$ and before termination of the random walk. It is easy to see that this number is not smaller than 1 (at least one visit at the initial step) and not greater than $(1-\alpha^2)^{-1}$ (there is a geometric number of returns, while $\alpha^2$ is the maximal possible return probability). Thus, it is sufficient to consider the variance of (II).

We will do this by using the local tree approximation. Choose $t:=t(n)$ as in Lemma~\ref{l:coupling}. Consider only $t$-neighborhood of $i$, denoted by $G_i^t$, and let $\tilde{\pi}^t_i$ be such that the contribution of (II) in (\ref{eq:decomposition}) is restricted only by paths in $G_i^t$:
\begin{align}
  \label{eq:decomposition2}
  \tilde{\pi}^t_i&=\frac{(1-\alpha)}{|{\cal S}|}\,\left(\sum_{s\in {\cal S}}\,\P_{(s)}(\mbox{$X_l=i$ for some $l\le T$, $X_0,\ldots,X_{l-1}\in G_i^t$})\right)\mathbb{E}_{(i)}\left[\sum_{l=0}^{\infty}1\{X_l=i\}\right]\\
  &=(I)\times (II)'\times (III),\nonumber
  \end{align}
Note that
\begin{equation}
\label{eq:tilda}\tilde{\pi}^t_i  \le  {\pi}_i,\end{equation}
because $\pi_i$ includes all paths of length $t$ plus some paths of lengths longer than $t$, which make loops in $G_i^t$ on the way to $i$, plus the paths that include a loop from $i$ back to $i$.
Thus, if we write 	$\pi_i = \tilde{\pi}^t_i+\delta^t_i$ with $\delta^t_i \ge 0$ then $\mathbb E(\delta^t)=o(n^{-1})$ due to (\ref{eq:tilda}) and Lemma~\ref{l:finite_neighborhood}.
It follows that
	\begin{align}
	\label{eq:varpii_bound}
	\mbox{Var}(\pi_i)& = \mbox{Var}(\tilde{\pi}^t_i)+\mbox{Var}(\delta^t)+2\E(\tilde{\pi}^t_i\delta^t) -2\E(\pi_i)\E(\delta^t)\\&\ge  \mbox{Var}(\tilde{\pi}^t_i)-2\E(\pi_i) o(n^{-1}) = \mbox{Var}(\tilde{\pi}^t_i)+ o\left([\E(\pi_i)]^2\right).
	\end{align}
	Therefore, it is sufficient to bound $\mbox{Var}(\tilde{\pi}^t_i)$ from below by a term of the order at least $[\mathbb E(\pi_i)]^2$. To this end, it follows from the same argument as after equation (\ref{eq:decomposition}), we need to analyze $\mbox{Var}((II)')$.
	
Conditioning on the last step before reaching $i$, we get:
  \begin{equation}\label{eq:II}(II)'=\sum_{j: j\; \mbox{\scriptsize offspring of } i}\,\frac{\alpha}{d_j}\sum_{s\in {\cal S}}\, \P_{(s)}(\mbox{$X_l=j$ for some $l\le T-1$, $j$ reached before $i$, $X_0,\ldots,X_{l-1}\in G_i^t$}).\end{equation}
Next, denote by $C_n$ the event that the $t$-neighborhood of $i$ coincides with the Galton-Watson tree $({\cal T}_i^t,\tau^t)$.
Conditioned on $C_n$, the terms in the external summation in $(II)'$ are independent. In particular, $\mbox{Var}((II)'|C_n)$ is a sum of three independent contributions: from the neighbors of $i$ in $\cal S$, ${\cal C}\backslash {\cal S}$ and $\bar{\cal C}$. We will lower bound $\mbox{Var}((II)'|C_n)$  by considering only the contribution of the neighbors of $i$ that are seed nodes. We number such neighbors as $i_1,i_2,\ldots,i_{N_0}$. Then we obtain
\begin{align*}
 &\quad \mbox{Var}((II)'|C_n)\\ &\ge \mbox{Var}\left(\sum_{j\in\{i_1,\ldots,i_{N_0}\}}
 \frac{\alpha}{d_{j}}\sum_{s\in {\cal S}}\mathbb P_{(s)}(\mbox{$X_l=j$ for some $l\le T-1$, $j$ reached before $i$, $X_0,\ldots,X_{l-1}\in G_i^t$})|C_n\right)\\
  &\quad:= \mbox{Var}\left(\sum_{j\in\{i_1,\ldots,i_{N_0}\}} Z_j|C_n\right)\\
  &\quad = \mathbb E(N_0|C_n)\text{Var}(Z_j|C_n) + \mbox{Var}({N_0}|C_n)(\mathbb E Z_j|C_n)^2.
  \end{align*}
	 Motivated by the above expression, we will evaluate the moments of $N_0$ given $C_n$. Recall that in the original graph, $N_0$ has $\mbox{Binomial}(k(n),q(n))$ distribution. Now, for $r>0$ and some $\epsilon<\xi/r$ we split $\E(N_0^r)$ as follows:
	\begin{equation}
	\label{eq:expectation}
	\E(N_0^r)=\E(N_0^r|C_n)\P(C_n)+E(N_0^r1\{N_0<n^\epsilon\}1\{\bar{C}_n\})+E(N_0^r1\{N_0>n^\epsilon\}1\{\bar{C}_n\}).
	\end{equation}
	By Lemma~\ref{l:coupling}, the second term in (\ref{eq:expectation}) is bounded from above by $n^{r\epsilon}\P(\bar{C}_n)=O(n^{-1+r\epsilon+o(1)})=o(k(n)q(n))$. The third term in (\ref{eq:expectation}) is bounded by $k^r(n)\P(N_0>n^\varepsilon)$. Using the bound from Theorem~2.21 in \cite{Hofstad}, we obtain that
	\begin{align*}
	k^r(n)\P(N_0>n^\varepsilon)&\le k^r(n)e^{-\frac{(n^\varepsilon-k(n)q(n))^2}{2(2k(n)q(n)/3+n^\varepsilon/3)}}\\&=k^r(n)O(e^{-n^\varepsilon/2})=o(k(n)q(n)).\end{align*} It follows that
		\begin{align*}\E(N_0|C_n)\P(C_n)&=k(n)q(n)(1+o(1)),\\
	\mbox{Var}(N_0|C_n)& = k(n)q(n)(1-q(n))(1+o(1)).
	\end{align*}
	From this and $q(n)=o(1)$, we conclude that for some $0<\gamma<1$ we have
	\begin{equation}\label{eq:varII'}\mbox{Var}((II)'|C_n)\ge \gamma k(n)q(n)\mathbb E(Z_j^2|C_n). \end{equation}
			Note that for every $j\in {\cal S}$ we have the trivial lower bound
	\[Z_j\ge \frac{\alpha}{d_j}\,\mathbb P_{(j)}(\mbox{$X_l=j$ for some $l\le T-1$, $j$ reached before $i$, $X_0,\ldots,X_{l-1}\in G_i^t$ })=\frac{\alpha}{d_j}.
\]
Further, recall that given $C_n$, $d_j \stackrel{d}{=} 1+\text{Poisson}(mq(n)+(n-m)q(n)).$ It follows that
\begin{align}
	\nonumber
    \mathbb E (Z_j^2|C_n) &\ge \alpha^2 \mathbb E \left( \frac{1}{d_j^2} \right)\ge \alpha^2 \left( \frac{1}{[\E( d_j)]^2} \right)\\
		\nonumber
    &=  \frac{\alpha^2}{(1 + mq(n) + (n-m)p(n))^2}\\
    \label{eq:z2}
		&\ge \frac{\alpha^2}{4n^2 q(n)^2},
\end{align}
where in the second inequality we used Jensen's inequality. From (\ref{eq:varII'}), (\ref{eq:z2}) it follows that
\[
    \text{Var}((II)'|C_n) \ge \frac{\gamma \alpha^2k(n)q(n)}{4n^2q(n)^2}.
  \]
Hence, since $(III)\ge 1$, from (\ref{eq:decomposition2}), we get
\[
  \text{Var}(\tilde{\pi}^t_i|C_n) \ge \frac{\gamma \alpha^2(1-\alpha)^2}{4k(n)q(n)n^2}.
\]
Finally, using that $\mathbb E(\pi_i)=\Theta(n^{-1})$ and $\lim_{n\to\infty}\P(C_n)=1$, we obtain
\begin{equation}\label{eq:tree}
\frac{\text{Var}(\tilde{\pi}^t_i)}{[\mathbb E(\pi_i)]^2}\ge \frac{\text{Var}(\tilde{\pi}^t_i|C_n)P(C_n)}{[\E({\pi}_i)]^2}=\Omega\left(\frac{1}{k(n)q(n)}\right),
\end{equation}
which, together with (\ref{eq:varpii_bound}), gives the result. \hfill $\Box$

\begin{remark}
It should be possible to relax the condition $k(n)=\omega(n^\xi)$ to $k(n)=\omega(1)$. For that, we need either a stronger coupling than in Lemma~\ref{l:coupling} or another way to evaluate $E(N_0|C_n)$ instead of (\ref{eq:expectation}).
\end{remark}

Let us now discuss some implications of Theorem~\ref{thm:nonconPPR}. Suppose, for example, that $q(n)=(1+a)p(n)$ for some $a>0$. The non-vanishing coefficient of variation means that $\pi_i$ has finite spreading around its mean. Then, in practice, if $a$ is small, we will not be able to distinguish many nodes in ${\cal C}$ from the nodes outside of $\cal C$, even in a very large network. We will provide an illustration for this scenario in Figures~\ref{fig:PPRk20},~\ref{fig:PPRk2} in Section~\ref{sec:localcluster}.

The necessary condition $k(n)q(n)\to\infty$ has the following very intuitive interpretation. Note that $k(n)q(n)$ is the average number of neighbors from $\cal S$ of a node in $\cal C$. Recall that each seed node (in $\cal S$) receives a certain large probability mass. When node $i$ has a finite average number of neighbors from $\cal S$, then their total contribution to $\pi_i$ is a finite random variable, so there is no concentration. Moreover, when $k(n)q(n)\to\infty$ then contributions of $\cal S$ to $\pi_i$ is a sum of asymptotically infinite number of terms, so the concentration should occur.

In the proof we used the coupling of the graph with a tree. Note that recent work \cite{Garavaglia2018LWC} allows one to pass the distribution of PageRank to the limit when the graph converges (possibly to a tree) in the `local weak convergence' sense. However, such convergence is defined only for sparse graphs, i.e., with asymptotically finite degrees, and does not apply to our
`medium dense' case (\ref{eq:semidense}). Also, Theorems~\ref{thm:L2conv}~and~\ref{thm:concentration_U}
are applicable to the dense regime when $p$ and $q$ do not depend on the size of the graph.

%
%

\section{Optimization with respect to the damping factor}
\label{sec:dampingopt}

In clustering applications, the performance of personalized PageRank is influenced by the choice of the parameter $\alpha.$ In \cite{ACL06}, the authors choose $\alpha$ as a function of the conductance of the desired smallest cut.
Typically, the conditions of \cite{ACL06} lead to the values of $\alpha$ very close to one.
In community detection applications, the probability of an error is a function of the difference between PageRank scores within and outside the community. In Theorem~\ref{thm:concentration_U}, we have identified a regime, where PPR $\pi$ is concentrated around its mean-field proxy $\bar \pi$. In such regime we can use the expressions for $\bar \pi_1$ and $\bar \pi_2$ from Section~\ref{sec:model} to find an \textit{optimal} parameter $\alpha$ that maximizes the difference between PageRank inside and outside community $\cal C$. Thus, the optimal $\alpha$ can be found as the solution to the following optimization problem:
$$
\alpha_{opt} = \arg \max_{\alpha} (\bar{\pi}_1(\alpha)-\bar{\pi}_2(\alpha)).
$$
Let us denote $\rho = \frac{q}{p}$ and $\beta = \frac{n-m}{n}.$ Then by (\ref{eq:barpi12}) we have
\begin{align*}
\bar{\pi}_1 - \bar{\pi}_2 &= {(\rho-1) (1-\alpha\beta)}\bar\pi_2\\
&= \frac{\alpha(1-\alpha)(\rho-1)(1-\alpha\beta)}{m\left(\alpha^2\beta(\rho-1) - \alpha (\rho\beta +\rho + \frac{\beta^2}{1-\beta}) + \rho+ \frac{\beta}{1-\beta}\right)}.
\end{align*}
The optimum $\alpha$ is such that $\left.\frac{d}{d\alpha}(\bar{\pi_1} - \bar{\pi_2})\right|_{\alpha = \alpha_{\mbox{\scriptsize opt}}} = 0.$ Thus, we find the optimum $\alpha$ as the solution of the following equation:
\begin{align*}
  \alpha^4 \beta^2(\rho-1) &-2\beta\left(\rho\beta + \rho + \frac{\beta^2}{1-\beta}\right) \alpha^3\\
	& 	+\alpha^2\left(3\beta(\rho + \frac{\beta}{1-\beta}) + (1+\beta) (\rho\beta + \rho+ \frac{\beta^2}{1-\beta}) -\beta(\rho-1)\right)\\& - 2\alpha(1+\beta)\left(\rho + \frac{\beta}{1-\beta}\right) + \rho + \frac{\beta}{1-\beta} = 0.
\end{align*}
With straightforward algebra, we can simplify the above equation as follows:
\[
(\alpha-1)^2\left(\alpha^2\beta^2(\rho-1) -2\alpha\beta\left(\rho+\frac{\beta}{1-\beta}\right) + \rho+\frac{\beta}{1-\beta} \right)= 0.
\]
Since $\alpha<1,$ the optimum is the solution of the following quadratic equation
\[
\alpha^2\beta^2(\rho-1) -2\alpha\beta\left(\rho+\frac{\beta}{1-\beta}\right) + \rho+\frac{\beta}{1-\beta}=0.
\]
The solutions to the above equation are
\[
\alpha^{*}_{i} = \frac{\rho-\beta(\rho-1) + (2i-1) \sqrt{\rho-\beta(\rho-1)}}{\beta(1-\beta)(\rho-1)},
\]
for $i=0,1.$ The solution corresponding to $i=1$ can be shown to be greater than 1 for any $\rho>1,\beta < 1$ and hence the only feasible solution is given by
\begin{equation}\label{opt_alpha}
  \alpha_{opt} = \min\left(1,\frac{\rho-\beta(\rho-1) - \sqrt{\rho-\beta(\rho-1)}}{\beta(1-\beta)(\rho-1)}\right).
\end{equation}

From the above equation, we can glean the following insight. Notice that if $x:= \rho-\beta(\rho-1) = (1-\beta)\rho+\beta,$ then after some elementary algebraic manipulations we have
\[
\alpha_{opt} = \frac{\sqrt{x}}{(1+\sqrt{x})(\rho-x)} = \frac{\sqrt{x}}{\beta(1+\sqrt{x})}.
\]
Thus, for a fixed $\rho$ or $\beta,$ $\alpha_{opt}$ is an increasing function of $x,$ while $x$ itself is an increasing (or decreasing) function of $\rho$ (or $\beta$). In other words, the more distinguishable the community is (larger $\rho$ or smaller $\beta$), the larger is the optimum $\alpha$. This conforms to the intuition that the RW starting from
the seed nodes should explore the graph more before termination when we have a denser or larger community.

\section{Numerical examples and implications for local graph\\ clustering}
\label{sec:localcluster}

The theoretical results of the two preceding sections have important implications for PPR based
local graph clustering. Our main theoretical result is that the parameter $k$ should scale linearly
and the parameter $m$ sufficiently fast with the size of the graph $n$ in order to ensure the concentration of PPR.
In practice, this means that the number of seed nodes should be significant to guarantee high quality clustering
results and the target community should not be too small.

As an aside, one could pose the following natural question: Can the sparsity-enforcing nature of the Approximate PPR (APPR) algorithm help
to avoid the leakage of probability mass to the nodes outside of the target community? Unfortunately, as we will demonstrate below in Subsection~\ref{subsec:APPR}, APPR suffers from the same non-concentration phenomenon as the original PPR.

\begin{figure}[!htb]
\centering
\includegraphics[scale=0.8]{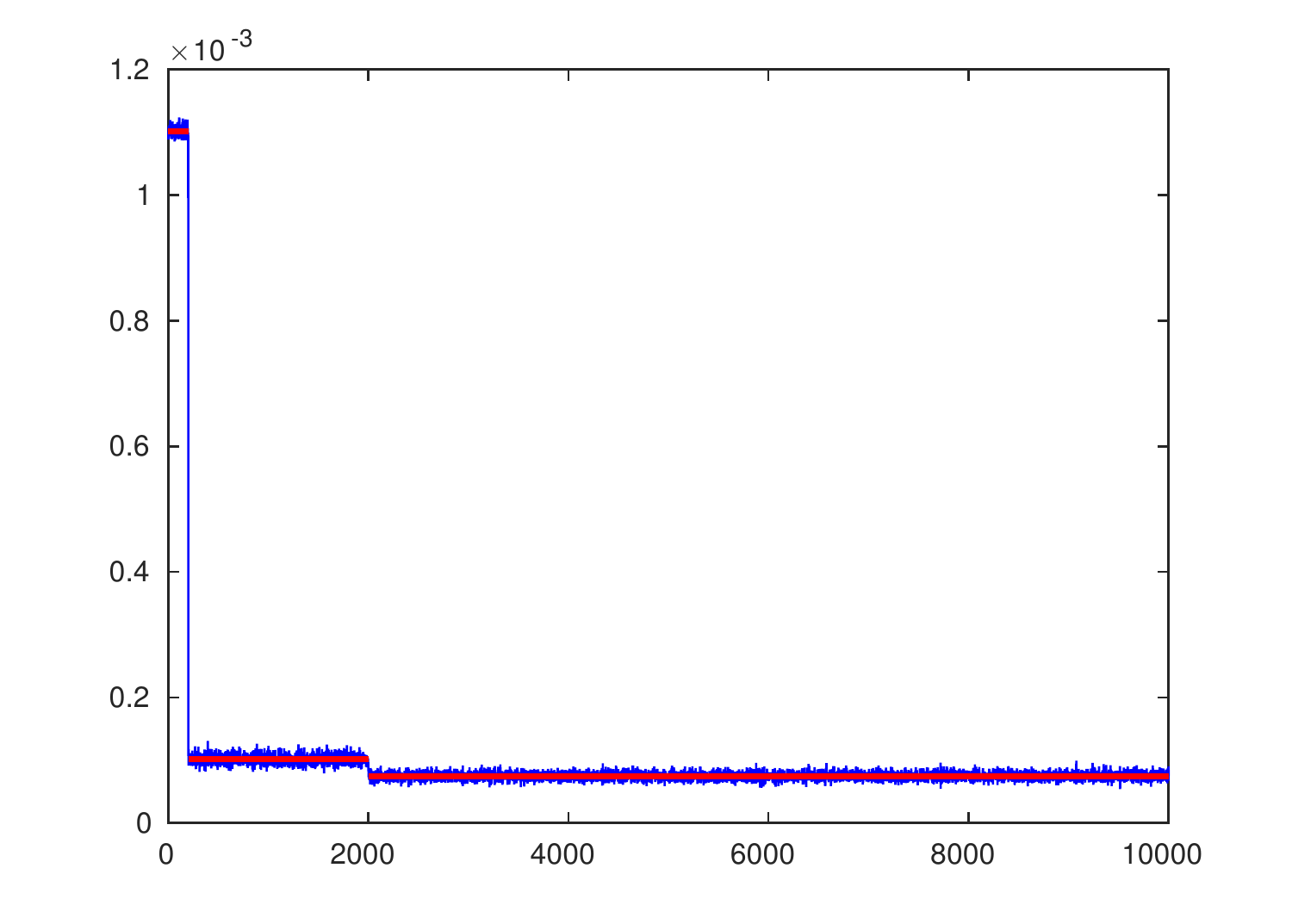}
\caption{PPR (blue) and its mean-field model (red) for k=200. On the $x$-axis are the indices of the nodes.
Nodes with indices $1,2,\ldots,2000$ belong to $\cal C$.}
\label{fig:PPRk200}
\end{figure}

For the purpose of illustration let us consider a specific numerical example.
We take $n=10000$, $m=2000$, and the edge probabilities as follows:
\begin{equation}
\label{eq:pqsetting}
p(n)=\frac{5\,\log^2(n)}{n}, \quad
q(n)=\frac{10\,\log^2(n)}{n},
\end{equation}
We first consider the case $\alpha=0.8$. If we set $k=200$, we observe a reasonably good concentration (see Figure~\ref{fig:PPRk200})
even though the values of $p(n)$ and $q(n)$ set by (\ref{eq:pqsetting}) imply very slow convergence with the rate
$1/\sqrt{\log(n)}$, according to (\ref{eq:bound_final}).
We can also calculate the percentage of nodes wrongly assigned to the community $\mathcal C$
according to the rank of PPR, which we denote by $\mathcal E$ defined below
\[
\mathcal E = \frac{|\overline{\mathcal C} \bigcap \widehat{\mathcal C}|}{|\mathcal C|},
\]
where $\mathcal C$ is the target community, $\overline{\mathcal C}$ is its compliment and
$\widehat{\mathcal C}$ is an algorithm output.
For the above chosen parameters we obtain $\mathcal E = 3.6\%$.

\begin{figure}[!htb]
\centering
\includegraphics[scale=0.8]{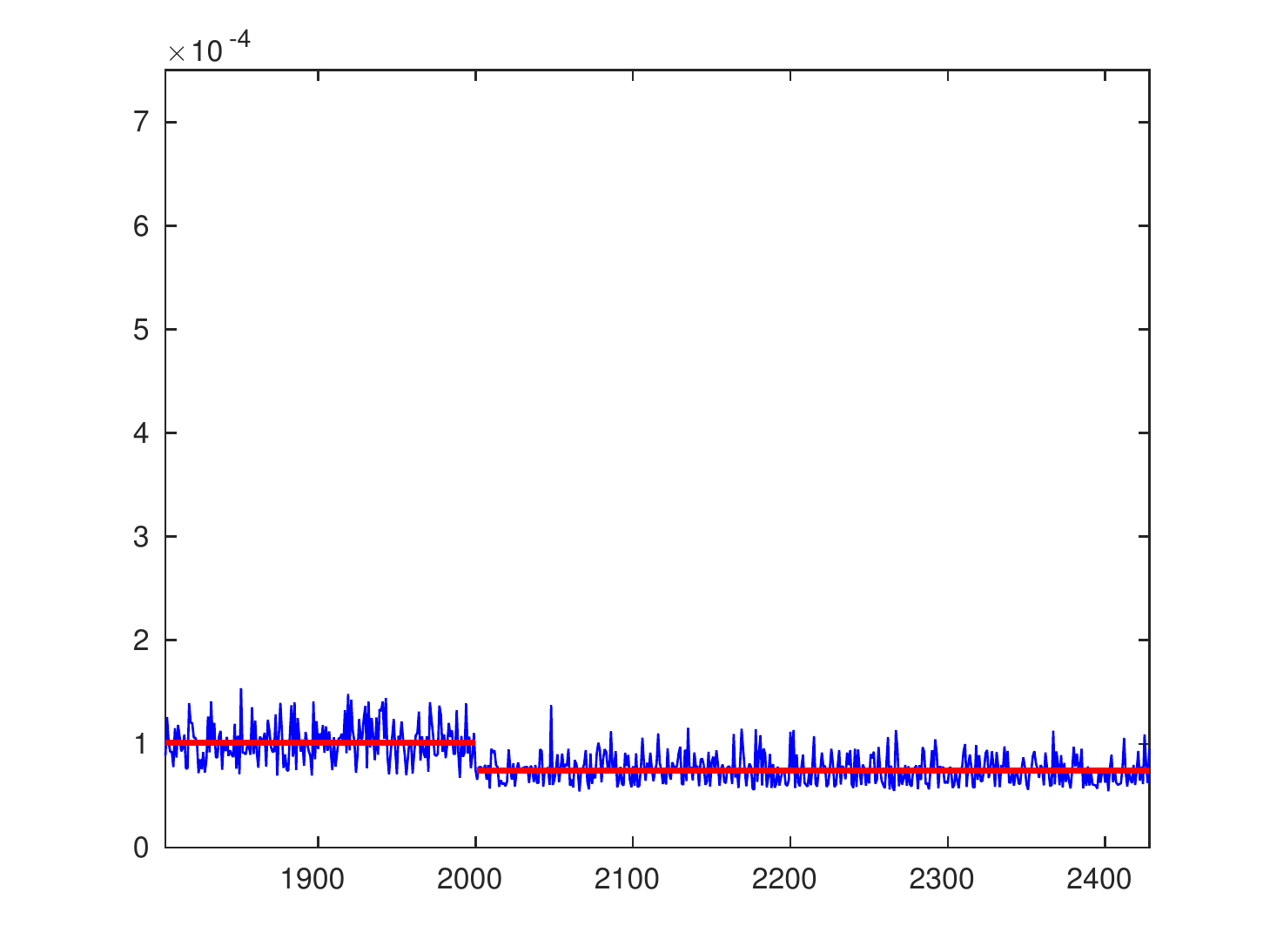}
\caption{PPR (blue) and its mean-field model (red) for k=20. On the $x$-axis are the indices of the nodes. Nodes with indices $1,2,\ldots,2000$ belong to $\cal C$.}
\label{fig:PPRk20}
\end{figure}

\begin{figure}[!htb]
\centering
\includegraphics[scale=0.8]{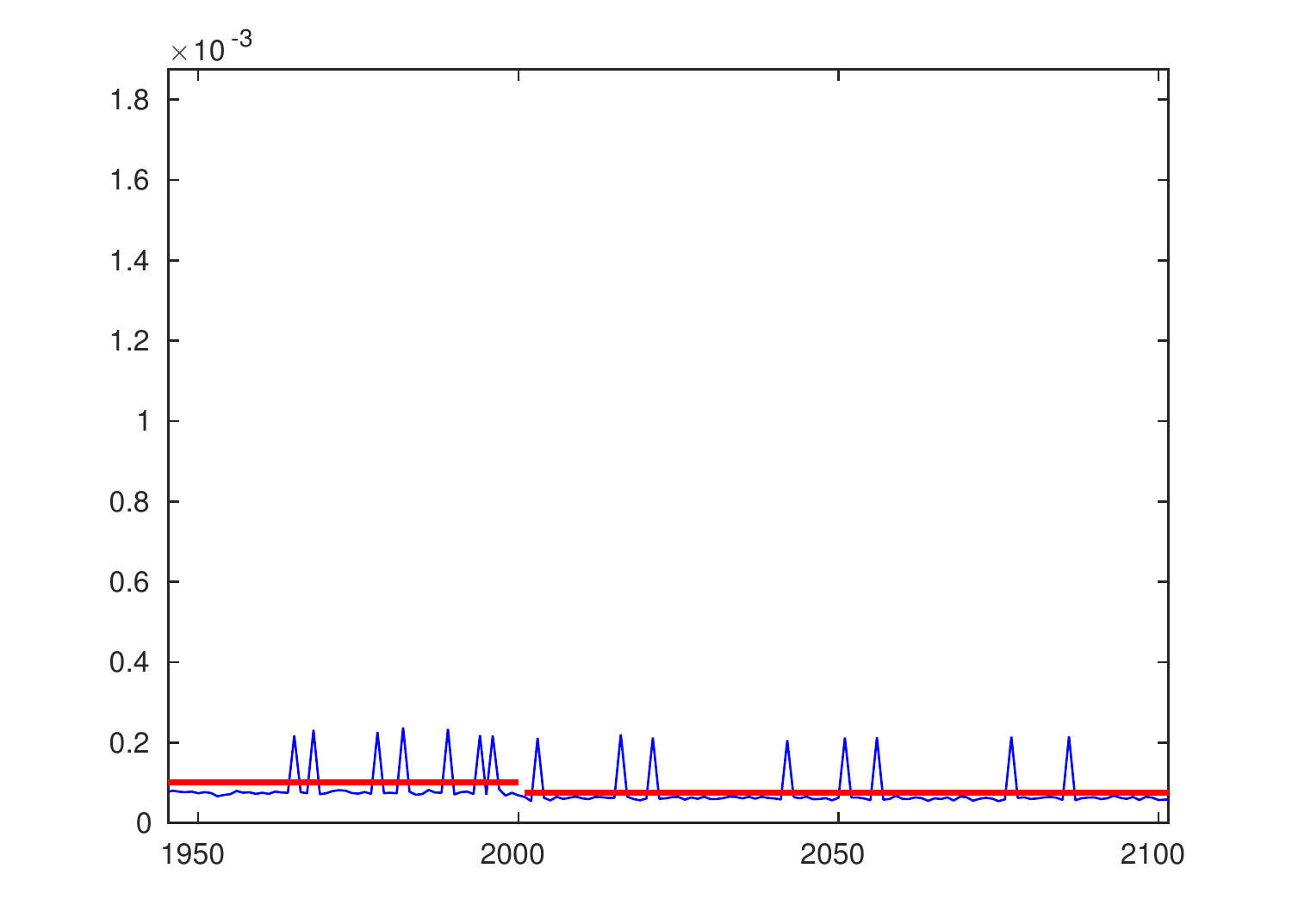}
\caption{PPR (blue) and its mean-field model (red) for k=2. On the $x$-axis are the indices of the nodes. Nodes with indices $1,2,\ldots,2000$ belong to $\cal C$.}
\label{fig:PPRk2}
\end{figure}

In the next experiment we decrease the number of seed nodes to $k=20$. Then the error increases by an order of magnitude
and becomes $\mathcal E = 44.2\%$, and we can observe the effect of non-concentration in Figure~\ref{fig:PPRk20}.
Curiously enough, if we decrease the number of seeds further to $k=2$, the error actually improves a bit to
$\mathcal E = 34.5\%$ but still remains very high. We can explain the slight decrease in the error by the
fact that most misclassified nodes are the neighbours of seed nodes. Notice that this is in accordance with
the proof of Theorem~\ref{thm:nonconPPR} where the neighbors that are seed nodes played a crucial role.
Indeed, the spikes in Figure~\ref{fig:PPRk2} correspond to the neighbours of the seed nodes. Thus, if we decrease
the number of seed nodes, we also subdue the main source of errors. Of course, there is a fine trade off
and one cannot eliminate completely the strong effect from non-concentration.

\subsection{Optimum value of $\alpha$}

\begin{figure}[!htb]
\centering
\includegraphics[scale=0.8]{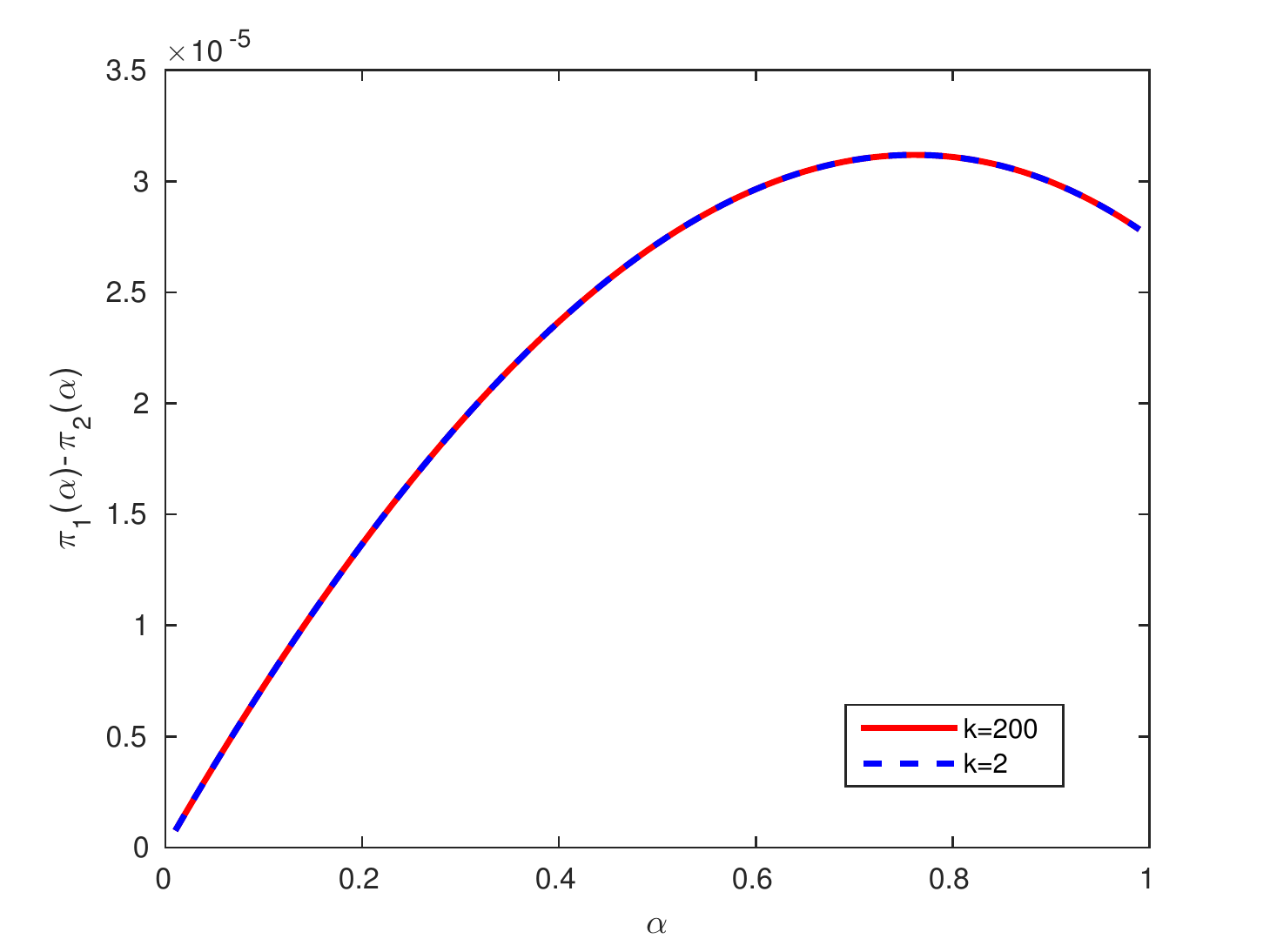}
\caption{Optimum value of the restart probability for $m=3000$.}
\label{fig:PPRdiffm3000}
\end{figure}

\begin{figure}[!htb]
\centering
\includegraphics[scale=0.8]{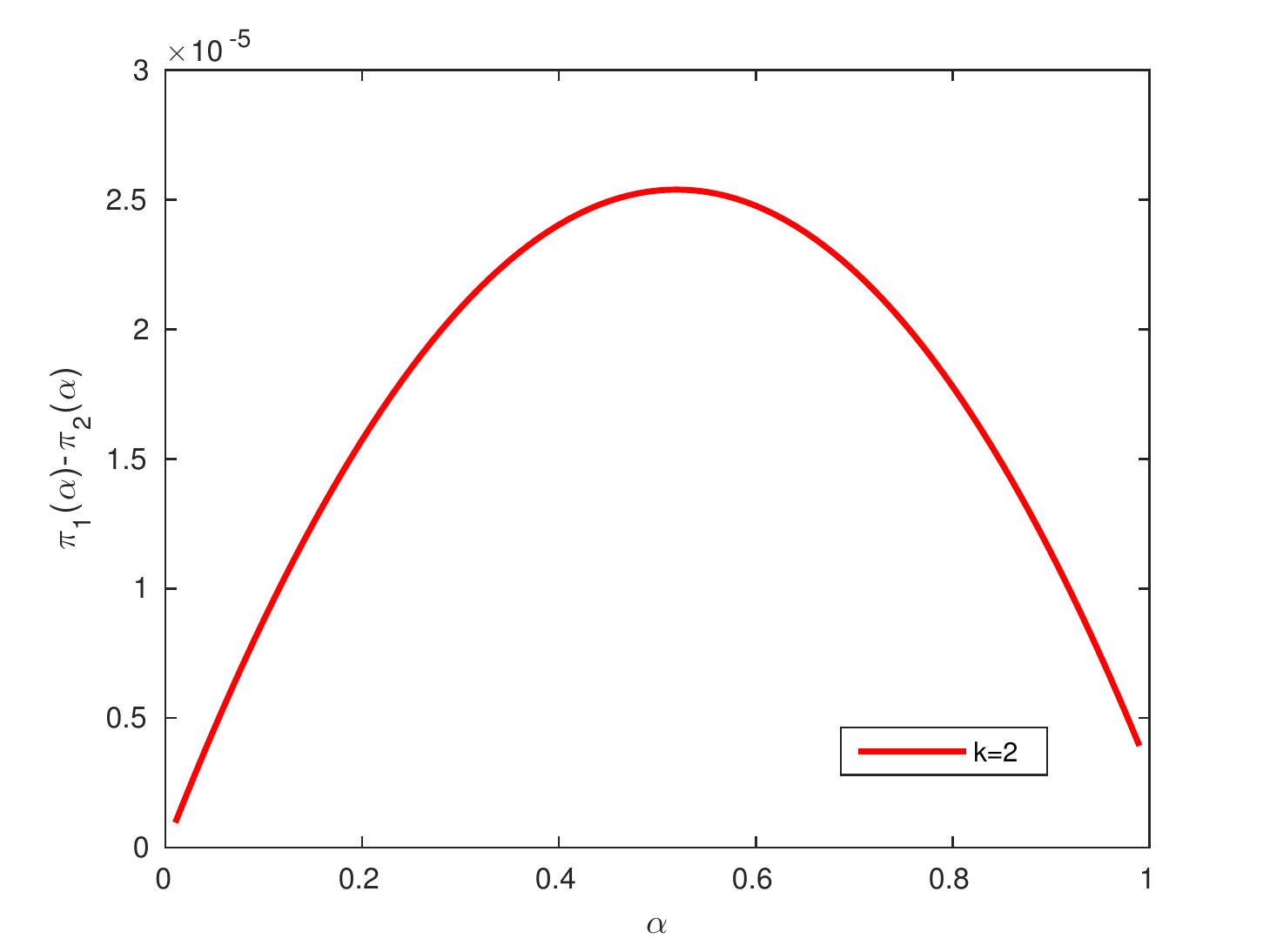}
\caption{Optimum value of the restart probability for $m=300$.}
\label{fig:PPRdiffm300}
\end{figure}

In this section we investigate numerically the dependence of the optimum $\alpha$ derived from the mean-field
version of PPR on the graph parameters.
In Figure~\ref{fig:PPRdiffm3000} we plot the difference $\bar{\pi}_1(\alpha)-\bar{\pi}_2(\alpha)$ as a
function of $\alpha$ for $n=10000$, $m=3000$, and $p$ and $q$ as in (\ref{eq:pqsetting}). We see that
the curves for $k=2$ and $k=200$ coincide. This is the case because $\bar{\pi}_1$ and $\bar{\pi}_2$ in
fact do not depend on $k$. It is interesting to observe that for reasonably large communities the optimum
value of $\alpha$ is quite close to the default value $0.85$ set by Google. Now, if we decrease the community
size from $3000$ to $300$, the optimum value of $\alpha$ decreases towards $0.5$ (see Figure~\ref{fig:PPRdiffm300}).
The decrease is expected, since to identify a smaller community, PPR needs shorter walks.
It might not be a coincidence that the optimum value of $\alpha$ decreased towards $0.5$, which was a value
recommended in \cite{ALP07,ALP08} by some other considerations.

\subsection{Non-concentration of Approximate Personalized PageRank}
\label{subsec:APPR}

The Approximate Personalized PageRank (APPR) algorithm in \cite{ACL06} is used to find a set of nodes $S$ with a given target conductance $\phi.$ In order to have a set with conductance at most $\phi,$ we need to choose the parameters of the algorithm $\alpha,\epsilon$ \cite{ACL06} such that $1- \alpha = \frac{\phi^2}{225\log(100\sqrt{|E|})}$ and $\epsilon = \frac{2^{-b} }{48B},$ where $B = \lceil{\log_2 (|E|)}\rceil$ and $ b \in [1,B]$ (Note: In our numerical experiments we take $b=13;$ larger values of $b$ decrease $\epsilon$ and thus increase the time to convergence of APPR, without significant gain in performance.)

For the graph parameter values considered in this section
the `mean form' conductance is $\overline\phi(\mathcal C) = 0.66$, see equation (\ref{eq:conduc}) below.
The conditions of \cite{ACL06} give $\alpha  = 0.999.$
Using these values we run a clustering algorithm based on the exact PPR where nodes are ranked according to their pagerank scores and the output community is the set of first $m$ nodes. We also run the APPR algorithm \cite{ACL06} with $\epsilon = 10^{-7}.$ The algorithm is quoted here for the sake of completeness in Algorithm~\ref{algo1}. Note that the final step of the clustering algorithm based on the approximate PageRank is to perform a ``sweep operation'' on the nodes ordered in decreasing order of a ranking function on the nodes. The ranking function proposed in \cite{ACL06} is the values of APPR divided by the node degrees. In our simulations we also investigate the performance of the algorithm where the ranking function is the approximate PageRank without this degree scaling. The two cases are denoted as `with degree scaling' and `without degree scaling' respectively. Finally, if the output of the sweep has more nodes than $m,$ we take the first $m$ nodes (i.e., nodes with largest values of the ranking function).

We summarize the values of the error ${\cal E}$ in Tables \ref{table1} and \ref{table2}. In Table \ref{table1}, we choose $\alpha = 0.85$ (initially used by Google webranking \cite{Abpr}). In Table \ref{table2}, we use the values computed based on the formulae in \cite{ACL06}, but with $\alpha$ chosen to be 0.99 (the algorithm does not converge for $\alpha= 0.999$).
\begin{center}
\begin{tabular}{ |c|c|c| }
 \hline
 $\alpha = 0.85,$
 $\epsilon = 10^{-8}$ & without degree scaling & with degree scaling \\
 \hline
 PPR & 0.35 & - \\
 \hline
 APPR & 0.49 & 0.724545 \\
 \hline
\end{tabular}
\captionof{table}{The error ${\cal E}$ of Approximate and Exact PPRs for $\alpha = 0.85$}\label{table1}
\end{center}

\begin{center}
\begin{tabular}{ |c|c|c| }
 \hline
 $\alpha = 0.99,$
 $\epsilon = 10^{-7}$ & without degree scaling & with degree scaling \\
 \hline
 PPR & 0.044 & - \\
 \hline
 APPR & 0.069 & 0.72 \\
 \hline
\end{tabular}
\captionof{table}{The error ${\cal E}$ of Approximate and Exact PPRs for $\alpha = 0.99$}\label{table2}
\end{center}

\begin{figure}[!htb]
\centering
\includegraphics[scale=0.8]{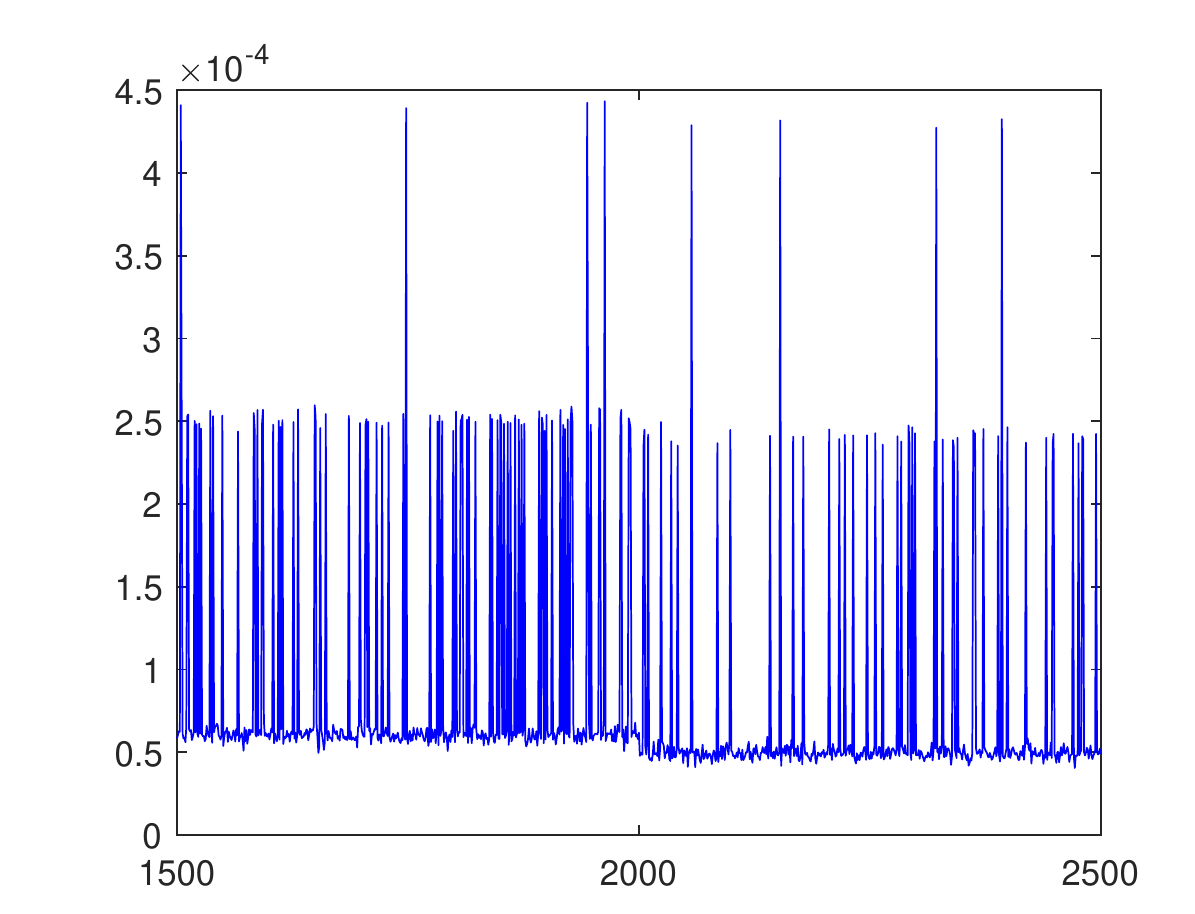}
\caption{APPR for $k = 2$ and $\alpha = 0.85$.}
\label{fig:p85}
\end{figure}

We make the following observations from our simulations. It is clear that APPR is also impacted by the same non-concentration phenomenon as the exact PPR. This is demonstrated in Figure~\ref{fig:p85} where we plot a realisation of APPR for $k=2$ and $\alpha = 0.85.$
The spikes again correspond to the neighbours of the seed nodes.

When $\alpha = 0.99$ the PPR solution is very close to the stationary distribution of a standard RW which is proportional to degrees. Hence we get almost perfect reconstruction in this case, since the expected degrees of the nodes can be used to cluster the graph nodes efficiently. 

This observation stems from the fact that we chose $m$ that grows linearly with $n$ and hence the degrees of the nodes inside the community are sufficiently different from the degrees of the nodes outside it. A more interesting scenario is a situation where $m = o(n).$ In this case, asymptotically the degrees of nodes outside the community and inside the community converge to the same value, making it impossible to detect the community only using the node degrees.

Let us take $m = 200, n = 10000, p = 5\frac{\log^2(n)}{n}$ and $q = 10\frac{\log^2(n)}{n}.$ By simply ranking by degrees and choosing the first $m$ nodes, we get error $\mathcal E = 0.935.$ Notice that if we do random guessing we get an error value $\mathcal E = 1-\frac{m}{n} = 0.98.$ Hence ranking based on degrees is almost as bad as random guessing! But using PPR we can get an error of $0.77$ with $\alpha = 0.7$ just with 20 seed nodes.

\begin{algorithm}
\caption{Clustering Algorithm using APPR}\label{algo1}
\begin{algorithmic}[1]

  \State Compute approximate pagerank vector $\text{pr}(v,\alpha,\epsilon)$ as in \cite{ACL06}.
  \State Do sweep operation:

    \State Sort vertices in decreasing order of  $\frac{\text{pr}(v,\alpha,\epsilon)_i}{d_i}$
    for $1\le i\le N_p,$ where $N_p$ is the maximum size of the subgraph.
    \State For the recursive node-set $S_i = \{1,2,\ldots i  \}$ at step $i,$ let $\phi_i$ be the conductance.
    Then $S_{\text{out}} = \arg \min_{i \le N_p}\phi_i.$

\State Return $S_{\text{out}}$ if $\phi(S_{\text{out}}) < \phi.$
\end{algorithmic}
\end{algorithm}

\subsection{Minimum conductance set}
\label{subsec:cond}

Notice that the conductance of community $\mathcal C,$ denoted by $\textnormal{cond}(\mathcal C)$ is given by
\begin{align}
\textnormal{cond}(\mathcal C) &= \frac{|\delta E|}{\min(\textnormal{vol}(\mathcal C),
\textnormal{vol}(\overline{\mathcal C}))}\nonumber\\
&= \frac{\sum_{i=1}^m \sum_{j=m+1}^n A_{ij} }{\min\left(\sum_{i=1}^m d_i, \sum_{i=m+1}^n d_i \right)}\label{eq:commcond}.
\end{align}
By virtue of Bernstein's inequality applied to the numerator and the degree concentration lemma applied to the denominator,
we can see that
the conductance of the community $\mathcal C$ converges to and can be well-approximated for a finite $n$
by $\overline \phi$ given below.
\begin{equation}\label{eq:conduc}
\overline \phi(\mathcal C) = \frac{\kappa(1-\kappa) p}{\min((\kappa)^2 q,(1-\kappa)^2 p)+ \kappa(1-\kappa)p},
\end{equation}
where $\kappa:= \frac{m}{n}.$

In \cite{CKL17} and \cite{ZLM13} it has been observed that the graph conductance has significant limitations
as a criterion for graph clustering. We show that in our random graph model the minimization of the graph
conductance does not lead to the determination of the natural cluster.

Now, suppose we are looking for a set $\mathcal A$ with minimal conductance, so we want to assign fraction $\gamma$ of nodes to $\mathcal A$. Assume first that an edge between any two nodes is present with equal probability $p$. Then we have a very simple expression for the `mean' conductance of $\mathcal A$:
\begin{equation}\label{eq:conducS}
\overline \phi(\mathcal A) = \frac{\gamma(1-\gamma) p}{\min(\gamma^2 p,(1-\gamma)^2 p)+ \gamma(1-\gamma)p}.
\end{equation}
 It is easy to see  that $\overline \phi(\mathcal A)$ is minimized when $\gamma=1/2$, so that in the denominator we get $\gamma^2=(1-\gamma)^2$. Now, assume that $q=(1+c)p$, and fraction $\kappa$ of nodes are in a hidden community ${\cal C}$. For simplicity of understanding, consider the case when $\gamma>\kappa$, and ${\cal C} \subseteq S $. Then in (\ref{eq:conducS}) only one term in the denominator will change, namely, it will increase by $\kappa^2c$:
\[
\overline \phi(S) = \frac{\gamma(1-\gamma) p}{\min((\gamma^2 p+\kappa^2\,c),(1-\gamma)^2 p)+ \gamma(1-\gamma)p}.
\]
Clearly, the equality $\gamma^2 p+\kappa^2\,c=(1-\gamma)^2$ will now hold for $\gamma<1/2$, so the set of minimal conductance will reduce. However, note that the value of $\gamma$, which minimizes conductance, is a continuous function of $c$ and $\kappa$. For example, when $c$ and/or $\kappa$ are small, the conductance will be minimized on a set $S$ that contains almost 1/2 of the nodes. This explains why the sets with minimal conductance, which we find in our experiments according to Algorithm~\ref{algo1},
are typically much larger than ${\cal C}$.

Specifically, the conductance of the community returned by exact PPR using only the first $m$ largest elements is $0.6748.$
The output conductance of the set returned by the sweep algorithm for APPR is $0.0024.$ However this set is much larger than the target community (its size is 4686). But when this set is truncated to 2000, its conductance becomes $0.6809.$

We would like to mention that if one considers the conductance values of communities of size $m$ (like in \cite{Jetal15,Letal09}),
then such `restricted' conductance is minimized on the natural community ${\cal C}$, given the community ${\cal C}$ is
neither too large nor too dense.

Thus, in the context of PPR based local graph clustering,
the size of the community (if available) can provide a better guidance than the conductance.

\section{Conclusions and future research}
\label{sec:conclusions}

We analysed a mean-field model of Personalized PageRank on the Erd\H{o}s-R\'{e}nyi random graph
containing a denser planted Erd\H{o}s-R\'{e}nyi subgraph. We also studied the
optimization of the damping factor, the only parameter in Personalized PageRank.
Our main conclusion is that PPR concentrates in the regime when the community size
scales linearly and the number of seed nodes scales sufficiently fast with the size of the graph.
We also identify the regime where concentration does not occur.
We have also demonstrated that the truncation of APPR does not mitigate the non-concentration of PPR.
The main reason for non-concentration of PPR and APPR is the significant leakage of probability mass
via the neighbours of the seed nodes.
This raises concerns about obtaining high quality local clustering when the number of seed
nodes is small. Of course, we have studied a very particular model of a network with
community structure. At the same time this model appears to be a very natural benchmark
for local graph clustering algorithms. Our concerns
complement the limitations of PPR based clustering discussed in \cite{CKL17}
and \cite{ZLM13}. As in \cite{CKL17,ZLM13,Jetal15}, we also note that the plain conductance
might not be the best criterion for local graph clustering.

From \cite{BWX15,HWX16} we know that recovering a hidden community is easy
and can be done by light complexity algorithms if the size of the community
scales linearly with the size of the graph and this is possible even without
using the seed nodes. As our analysis indicates, there are concerns about
the applicability of PPR based method in the regime with sublinear scaling
of the number of seed nodes. In contrast, belief propagation based algorithms can achieve good detection performance even with a few number of seeds; however, they require good quality seeds and, unlike PPR, they require the
knowledge of the graph parameters \cite{Ketal17}. Possibly, a combination of these ideas is needed to overcome the limitations of both PPR and belief propagation algorithms.

One more interesting research direction is the extension of the present
results to the setting of multiplex networks \cite{DMR16,Hetal13} when several
networks represent one actual underlying phenomenon. We expect that using
several instances of the same network will significantly improve
concentration, and hence the performance, of the PPR based clustering methods.

\section*{Acknowledgements}
This work was partly funded by the French Government (National Research Agency, ANR) through the ``Investments for the Future'' Program reference \#ANR-11-LABX-0031-01, Inria - IIT Bombay joint team (grant IFC/DST-Inria-2016-01/448)
and by EU COST Project COSTNET (CA15109).

\begin{table}
\caption{Notation}
\label{tab:notation}       
\begin{tabular}{ll}
\hline\noalign{\smallskip}
Symbol & Meaning \\
\noalign{\smallskip}\hline\noalign{\smallskip}
$V$ & set of nodes \\
$\mathcal C$ & planted subgraph (community) \\
$\mathcal S$ & set of seed nodes \\
$n$ & size of the graph \\
$m$ & size of the planted subgraph \\
$k$ & number of seed nodes \\
$p$ & probability of edge in the graph \\
$q$ & probability of edge in the subgraph \\
$\onevec_n$ & vector of ones of dimension $n$ \\
$\mat{J}_{m,n}$ & matrix of ones of dimension $m$-by-$n$ \\
$\zerovec_n$ & vector of zeros of dimension $n$ \\
$A$ & adjacency matrix \\
$d=A\onevec_n$ & vector of nodes' degrees \\
$D=diag\{d\}$ & diagonal matrix of nodes' degrees \\
$P=D^{-1}A$ & transition probability matrix \\
$\pi$ & Personalized PageRank \\
$\alpha$ & damping factor \\
$\bar{A}$ & expected adjacency matrix \\
$\bar{d}=\bar{A}\onevec_n$ & vector of expected nodes' degrees \\
$\bar{D}=diag\{\bar{d}\}$ & diagonal matrix of expected nodes' degrees \\
$\bar{P}=\bar{D}^{-1}\bar{A}$ & mean-field transition probability matrix \\
$\nu$ & personalization vector or restart distribution \\
$\bar{\pi}$ & mean-field Personalized PageRank \\
$\bar{\pi}_0$ & mean-field Personalized PageRank of a seed node \\
$\bar{\pi}_1$ & mean-field Personalized PageRank of a subgraph node \\
$\bar{\pi}_2$ & mean-field Personalized PageRank of a node outside the subgraph \\
$\cal E$& percentage of nodes in $\cal C$ that are misclassified by the algorithm\\
$f(n)=\omega (g(n))$& $f$ dominates $g$ asymptotically\\
$f(n)=\Omega (g(n))$& $f$ is bounded below by $g$ asymptotically\\
$f(n)=\Theta (g(n))$& $f$ is bounded both above and below by $g$ asymptotically\\
\noalign{\smallskip}\hline
\end{tabular}
\end{table}



\end{document}